\documentclass{article}
\usepackage{spconf,amsmath,graphicx,citesort}
\usepackage{amsmath,graphicx}
\usepackage{mathrsfs}
\usepackage{amsfonts}
\usepackage{color, soul}
\usepackage{ctable}
\usepackage{multirow}
\usepackage{xspace}
\usepackage{citesort}
\usepackage{pifont}

\newcommand{\cmark}{\ding{51}}%
\newcommand{\xmark}{\ding{55}}%

\title{Microsoft Speaker Diarization System for the VoxCeleb Speaker Recognition Challenge 2020}

\name{\begin{tabular}{c}Xiong Xiao, Naoyuki Kanda, Zhuo Chen, Tianyan Zhou, Takuya Yoshioka \\Sanyuan Chen, Yong Zhao, Gang Liu, Yu Wu, Jian Wu, Shujie Liu, Jinyu Li, Yifan Gong\end{tabular}}
\address{Microsoft, USA}
\begin{document}
\ninept
\maketitle
\begin{abstract}
This paper describes the Microsoft speaker diarization system for monaural multi-talker recordings in the wild, evaluated at the diarization track of the VoxCeleb Speaker Recognition Challenge (VoxSRC) 2020. 
We will first explain our system design to address issues in handling real multi-talker recordings. 
We then present the details of the components, which include Res2Net-based speaker embedding extractor, conformer-based continuous speech separation with leakage filtering, and a modified DOVER (short for Diarization Output Voting Error Reduction) method for system fusion. We evaluate the systems with the data set provided by VoxSRC challenge 2020, which contains real-life multi-talker audio collected from YouTube. Our best system achieves 3.71\% and 6.23\% of the diarization error rate (DER) on development set and evaluation set, respectively, being ranked the 1st at the diarization track of the challenge. 
\end{abstract}
\section{Introduction}
\label{sec:intro}

Speaker diarization is the task of determining ``who spoke when" given a long audio signal \cite{tranter2006overview}.
It is an imporant component for audio analysis and has a wide range of application domains, such as broadcast news, meetings, and telephone conversations. It can also be used to improve automatic speech recognition  in multi-speaker conversation scenarios \cite{kanda2019guided,medennikov2020stc}.

There have been tremendous efforts for improving speaker diarization systems.
A speaker diarization system typically consists 
of several modules, including voice activity detection (VAD), speech segmentation, 
speaker embedding extraction, and speaker clustering.
Each module has been extensively studied for different purposes such as  
speaker embedding~\cite{dehak2010front,snyder2018x,variani2014deep,bredin2017tristounet,zhou2019cnn,wang2020speaker}
and speaker clustering~\cite{meignier2010lium,shum2013unsupervised,garcia2017speaker,wang2018speaker}. 
There has also been an international effort to find out
best practices that would work for a diverse set of recordings \cite{ryant2018first,ryant2019second}. 
Despite these advances, 
speaker diarization for real recordings still remains to be  challenging 
problem.

Difficulty of speaker diarization for real world recordings arises from (1) diversity of speaker characteristics and (2) adverse acoustic conditions, which often contain overlapping utterances (or simultaneously active speakers). 
Especially, the speech overlaps have sometimes been
excluded from the system design as well as
the evaluation metrics (e.g., \cite{shum2013unsupervised,garcia2017speaker,wang2018speaker})
due to the difficulty in handling them.
However, 
speech overlaps are frequently observed in real conversations.
The overlap ratio (the percentage of the time during which more than one person speaking) ranges from 10\% to 30\% for meetings \cite{ccetin2006analysis}, and it can become higher for daily conversations \cite{kanda2019guided,barker2018fifth,watanabe2020chime}. 
Recently proposed neural network-based diarization systems, such as 
end-to-end neural diarization \cite{fujita2019end,fujita2019endself}
and target-speaker voice activity detection \cite{medennikov2020stc}, 
were shown to be effective for overlapped speech. However, 
they have a limitation that the maximum number of
recognizable speakers is constrained by the number of output channels of the
neural networks. It is 
also 
not clear whether these model-based diarization techniques generalize to unseen conditions. 


\begin{figure}[t]
    \centering
    \includegraphics[width=\linewidth]{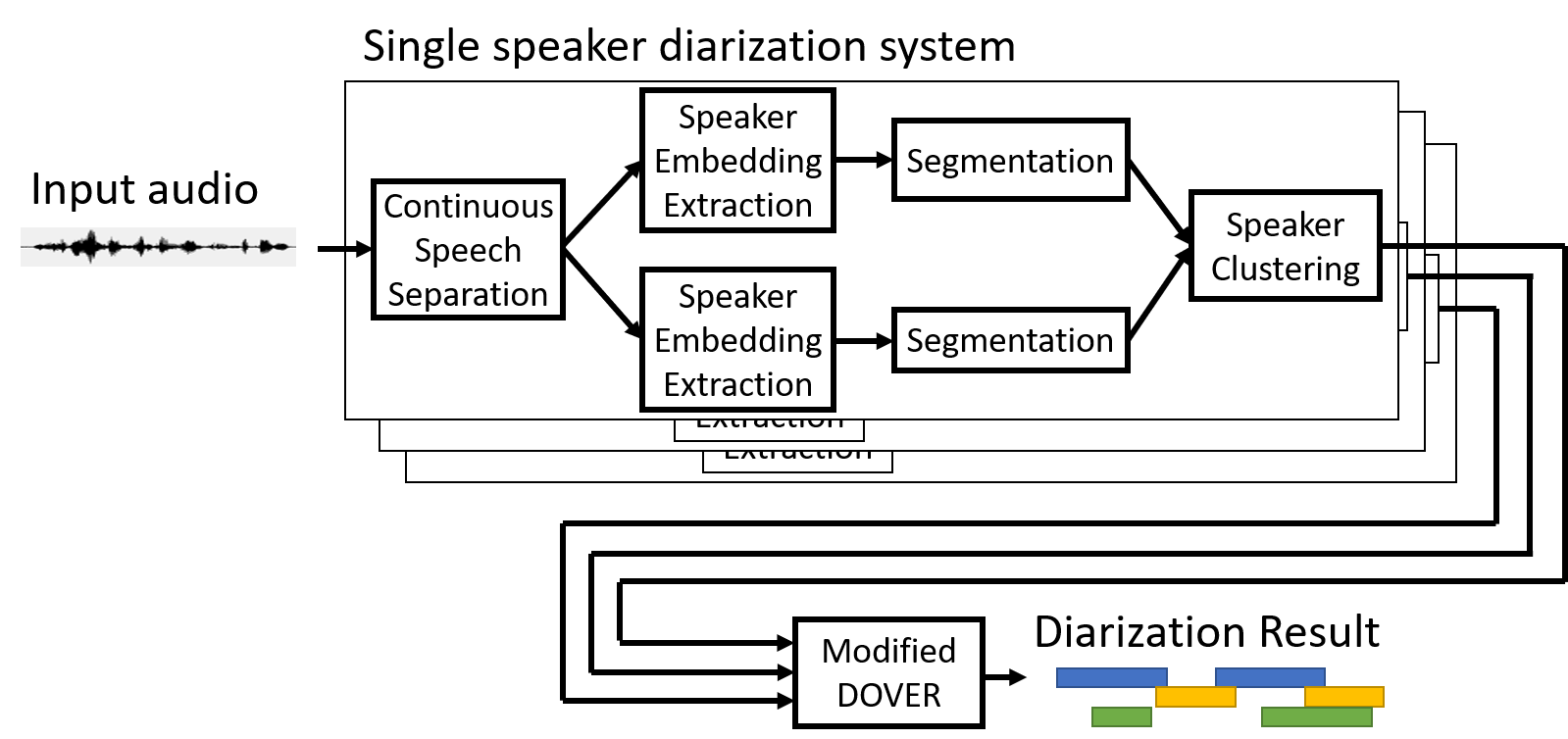}
    \caption{System Diagram}
    \label{fig:system_diagram}
\end{figure}

With this as a background,
we propose a speaker diarization system that
consists of continuous speech separation (CSS), speaker embedding extraction, segmentation,
speaker clustering,
and system fusion as shown in Fig.~\ref{fig:system_diagram}.
The prominent components of our system can be summarized as follows.
{
    \setlength{\leftmargini}{0pt}
    \begin{description}
    \setlength{\itemindent}{-5pt}
    \item {\bf Conformer-based CSS}\\
    We develop a highly optimized
    CSS system 
    based on the conformer network \cite{gulati2020conformer,chen2020continuous}
    to address the speaker overlap problem.
    \item {\bf Res2Net-based speaker embedding extractor}\\
     We incorporate Res2Net architecture \cite{gao2019res2net} with additive margin Softmax loss \cite{wang2018additive} to train our speaker embedding extractor, which enables highly accurate speaker clustering. 
    \item {\bf Speaker clustering with leakage filtering}\\
    A speaker clustering with a leakage filtering method is also proposed 
    to reduce the false alarm due to the residual noise and music in the output channels of the speech separation.
    The leakage filtering is essential to achieve a
    significant improvement by speech separation.
    \item {\bf System combination with a modified DOVER}\\
    We fuse the outputs of multiple diarization systems by using a novel voting-based algorithm, called modified DOVER, which is an extension of DOVER \cite{stolcke2019dover} to handle overlapped speech.
    %
\end{description}
}

Next section briefly describes the dataset used in the diarization track of VoxSRC challenge 2020.
Section \ref{sec:system} presents the proposed system.
Section \ref{sec:results} reports the evaluation results which are
the best result in the diarization track of VoxSRC 2020. The last section concludes the paper.

\section{VoxSRC Challenge 2020 Dataset}
\label{sec:dataset}

\begin{figure}[t]
    \centering
    \includegraphics[trim=2cm 0cm 0cm 0cm, width=0.52\textwidth]{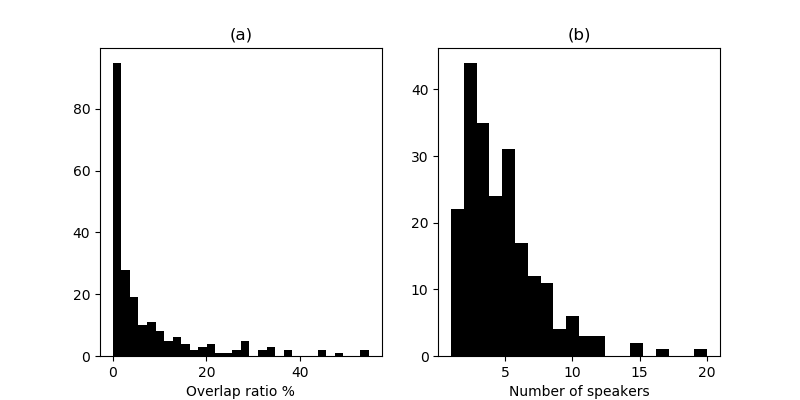}
    \caption{Statistics of recordings in development set, (a) histogram of overlap ratios; (b) histogram of number of speakers. }
    \label{fig:dev_set_statistics}
\end{figure}

At the VoxSRC Challenge 2020, the development data and the evaluation data 
were both monoaural multi-talker recordings provided by the organizer.
The dataset is obtained from YouTube videos, 
consisting of multi-speaker audio from both 
professionally edited videos as well as more casual conversational multi-speaker audio.
Throughout the audio, many artifacts are observed such as
background noise, music, laughter, applause, and singing voices, which make the speaker diarization challenging.
The audio also contains plenty amount of overlapped speech as shown in Fig.~\ref{fig:dev_set_statistics}(a).

The reference time information was provided only for development set,
and a participant can submit the result for evaluation  data only up to 5 times, once per day.
Therefore, we examined our system mainly based on the development set,
and only show the results of evaluation set for our system submissions (we submitted 4 systems in total).
The development set consists of 216 recordings (20.3 hours in total).
The number of speakers in one recording varies from 1 to 20 as shown in Fig.~\ref{fig:dev_set_statistics}(b), and the average overlap ratio of the recordings in the development set is 7.1\% according to the reference time information.
The evaluation set consists of 310 recordings (54.1 hours in total).

Evaluation was conducted based on the diarization error rate (DER) and the Jaccard error rate (JER) \cite{ryant2018first}.
The DER was used for the primary metric of the challenge, and we also tuned our system based on it. More information about the challenge data set and evaluation metrics can be found in \cite{Nagrani19}. 

\section{Proposed Speaker Diarization System}
\label{sec:system}

\subsection{Overview}
The proposed speaker diarization system is illustrated in Fig.~\ref{fig:system_diagram}. The input audio is first processed by the CSS module that separates potential overlapped speech in a blockwise manner. The output is always two separated channels. For regions with only one active speaker, one of the channel is supposed to contain the speech while the other is supposed to be empty. 
The separated channels are independently processed by speaker embedding extraction and segmentation modules in sequence. The segments from both channels are then pooled together and fed into the clustering module where they are grouped into speaker clusters. 
Finally, diarization outputs from multiple systems are fused by the modified DOVER. The details of each module will be explained in the following subsections. 

Note that our system can handle at most two simultaneous speakers by the design of the CSS module. 
In addition, in some configuration of our systems, we do not apply speech separation.
In such a case,
 the original mixed speaker signal is fed into the pipeline just by simply skipping the CSS module. 

\subsection{Conformer based CSS}

To handle the overlapped speech, the CSS framework~\cite{yoshioka2019css,chen2020continuous} is applied to each meeting due to its capability in handling arbitrary long sequence with various number of speakers in conversation. In this work, we assume the maximum number of simultaneously talking speakers is two. For each meeting, two channel outputs are estimated by separation module, where each channel only contains single active speaker. More detail can be found in \cite{yoshioka2019css,chen2020continuous}.

The frequency mask based approach were used for separation, where two masks were estimated for each frame of the input spectrogram. Following \cite{chen2020conformer}, we applied the conformer based separation network, which consists of 18 conformer encoder layers with 8 attention heads, 512 attention dimensions and 1024 FFN dimension. The model was trained with permutation invariant training objective with mean squared error between the magnitude spectrogram of the reference signal and masked mixture signal, with the mean and variance normalized spectrogram from mixture speech as input feature.

To train the network, we simulated 1500 hours of mixed training sample. For each sample, two clean speech utterances sampled from WSJ-1 and LibriSpeech \cite{panayotov2015librispeech} data sets were firstly convolved with room impulse response simulated with image method \cite{allen1979image}, then mixed with signal to noise ratio sampled between -5 $\sim$ 5 dB. We followed the mixing setup as in \cite{chen2020conformer}. Meanwhile, we randomly removed one mixing source in 10\% of training data to create single speaker mixtures.

\subsection{Res2Net-based Speaker Embedding Extractor}
\label{sec:embedding}
Res2Net \cite{gao2019res2net} structure was originally proposed for image classification. It introduced a new dimension, called scale, to improve ResNet model's representation power. In our previous work \cite{zhou2020resnext}, we investigated the effectiveness of Res2Net architecture for text-independent speaker verification (SV) task. Experimental results demonstrated that  increasing scale is more efficient than going deeper or wider. Res2Net model exhibits stronger capacity than conventional ResNet even with similar number of parameters. In addition, we verified Res2Net structure outperforms ResNet baseline for short utterances and mismatched scenarios due to its multi-scale feature representation ability, which could also benefit subsequent clustering.

Given these promising results,
in our diarization system, we 
adopted the Res2Net structure as our speaker embedding extractor. In order to further enhance speaker embedding's discrimination, we applied the additive margin Softmax (AM-Softmax) loss \cite{wang2018additive} as our training criterion. 
The integration of Res2Net structure and AM-Softmax loss brought us a state-of-the-art speaker embedding extractor.

As shown in Table~\ref{tab:results_sv}, we prepared three models with different configurations. 
The first and third models were trained with VoxCeleb1\&2, containing 7323 speakers in total (the VoxCeleb1-test part is excluded from training),
while the second model was trained with VoxCeleb2-dev, which contains 5994 speakers.
We also show 
the equal error rate (EER) and the minimum detection cost function (minDCF)
of all three systems on the standard VoxCeleb1 test set.

\begin{table}[htb]
\setlength{\tabcolsep}{3.5pt}
	\caption{Evaluation results with different model structures. Notation for model: (w: base width; s: scale). EER (\%) and minDCF (p\_target=0.05) are reported on VoxCeleb1 test set.}
	\label{tab:results_sv}
	\centering
    \begin{tabular}{c l c c c c}
    	\hline
    	\multirow{2}{*}{\textbf{ID}}&
    	\multirow{ 2}{*}{\textbf{Model}} & 
    	\multirow{ 2}{*}{\textbf{\#spks}} &
    	\multirow{ 2}{*}{\textbf{Loss}} &
    	\multicolumn{2}{c}{\textbf{VoxCeleb1-test}} \\  
    	\cmidrule{5-6}
    	& & & & \textbf{EER} & \textbf{minDCF} \\ \hline
        $E_1$ & Res2Net23-26w8s & 7323	& Softmax &	1.16 &	0.0737 \\ \hline
        $E_2$ & Res2Net50-26w8s & 5994	& AM-Softmax &  0.90 & 	0.0509	\\ \hline
        $E_3$ & Res2Net50-26w8s & 7323	& AM-Softmax &	0.83  &	0.0473  \\ \hline
    \end{tabular} 
\end{table}

\subsection{AHC-based Segmentation}
\label{sec:segmentation}

Speaker segmentation module segments the continuous audio into multiple short segments so that each segment contains only one speaker. 
This is performed in two steps. 
First, voice activity detection (VAD) is applied to the input audio to extract each continual speech region where at least one person is active. 
Then, each region is further decomposed into the speaker-homogeneous segments by means of agglomerative hierarchical clustering (AHC).
To do so, speaker embedding vectors are extracted at the rate of 12.5 Hz (i.e., one vector for every 80 ms). Each pair of two consecutive vectors is grouped to form an initial set of segments. For every neighboring segment pair, 
the degree of proximity between the two
segments is estimated in the embedding space. The closest
pair is then merged to form a new longer segment. The proximity
is defined as the cosine similarity between the mean embedding of the two segments.
This process is repeated until the cosine similarity becomes less than 
a pre-determined threshold. 

\subsection{Speaker Clustering with Leakage Filtering}
\subsubsection{AHC-based speaker clustering}
AHC is used to group speech segments into clusters. A high stopping threshold is used in the AHC to ensure high speaker purity of the clusters. As a result, the number of clusters is usually much larger than the number of actual speakers. Speaker clusters are chosen from the clusters according to a duration criterion. Specifically, only those clusters that are longer than a predefined minimum speaker duration are considered as a valid speaker cluster. After that, a speaker embedding centroid vector is obtained for each speaker cluster, and the rest of the clusters are assigned to one of the speaker clusters via cosine similarity. The motivation of the above strategy is to increase the chance of the main speakers being diarized correctly while sacrificing the performance on minor speakers. To avoid assigning minor speakers' clusters, which failed to be treated as a valid speaker cluster, to other speaker clusters, a SV step is introduced. If the similarity between a cluster to be assigned and its most similar speaker cluster is lower than an SV threshold, the cluster will not be assigned to the speaker cluster but treated as unassigned cluster. Currently, all unassigned clusters in a session are treated as one single cluster during DER and JER computation. In practice, special treatment of these unassigned clusters are required which usually depends on the application of speaker diarization. The AHC stopping threshold, minimum speaker duration, and SV threshold are tuned on the development set and set to 0.55, 2.5 seconds, and 0.0, respectively. 

\subsubsection{Leakage filtering} 
The CSS module occasionally produces residual noises. For example, when a single speaker is speaking with background music, 
one of the two separated signal may contain the active speaker's voice, while the other one may contains residual music. The VAD may incorrectly tag these residual noise/music as speech. One solution to this problem may be to use automatic speech recognition (ASR) as a VAD, but this makes the system language dependent and introduces high computational cost and latency. To handling this issue, we introduced a speaker embedding based segment filtering step. 

The diarization system is run on both 
with and without the CSS module independently.
From the diarization output of the system without the CSS module, 
we obtain a set of speaker clusters and we assume that the centroids of these clusters contain all the speakers' signature in the audio. 
These centroids are used to filter the speech segments from the system with the CSS module. 
Specifically, if the maximum cosine similarity of a segment to the centroids is below a predefined threshold, the segment will be removed from the diarization output. We found that this approach is able to reduce the VAD errors on separated channels significantly. The filtering threshold is set to 0.2 which is tuned from development set. 

 \begin{figure}[t]
  \centering
  \includegraphics[width=\linewidth]{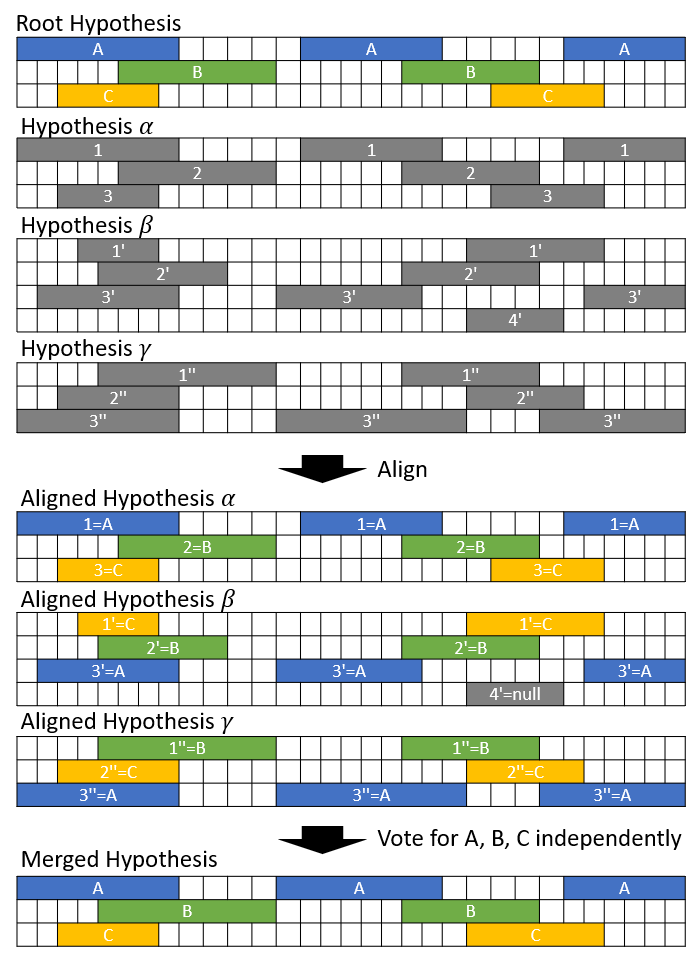}
  \caption{Example of the modified DOVER on three hypotheses. In the root hypothesis, there are three speakers called A, B, and C. Three hypotheses $\alpha$, $\beta$, and $\gamma$ are merged along with the root hypothesis.}
  \label{fig:modified_dover}
\end{figure}

\begin{table*}[t]
  \caption{The DER (\%) and JER (\%) of the proposed speaker diarization system. Speaker clustering ``$C_1$'' uses stopping threshold of 0.6 and minimum speaker duration of 4s, while speaker clustering ``$C_2$'' uses 0.55 and 2.5s, respectively. }
  \label{table:results}
  \centering
  {
  \begin{tabular}{cc|cccc|cc|cc}
    \hline
System & Submission & Speaker  & Speaker & Speech & Leakage &\multicolumn{2}{c|}{Dev} & \multicolumn{2}{c}{Test} \\
&&embedding & clustering & separation & filtering & DER (\%) & JER (\%) & DER (\%) & JER (\%) \\ \hline
Baseline &-   & - & - & - & - & - & - & 21.75 & 51.89 \\ \hline
1 & 1 & $E_1$ & $C_1$ & \xmark & \xmark & 5.63 & 25.65 & 8.87 & 21.08 \\
2 & 2 & $E_2$ & $C_1$ & \xmark & \xmark & 5.06 & 24.47 & 8.54 & 20.58 \\
3 & - & $E_3$ & $C_1$ & \xmark & \xmark & 5.04 & 23.97 & -    & - \\
4 & - & $E_3$ & $C_1$ & \cmark & \xmark & 4.91 & 23.33 & -& - \\
5 & 3 & $E_3$ & $C_1$ & \cmark & \cmark (Sys. 3) & 3.89 & 23.02 & 8.08 & 17.78 \\
6 & - & $E_3$ & $C_2$ & \xmark & \xmark          & 4.91 & 19.90 & - & - \\
7 & - & $E_3$ & $C_1$ & \cmark & \cmark (Sys. 6) & 3.80 & 18.69 & - & - \\ \hline
8 & 4 & \multicolumn{4}{c|}{Fusion of System 1, 2, 3, 7}          & 3.71 & 18.74 & 6.23 & 21.52 \\
    \hline
  \end{tabular}
  }
  \vspace{-3mm}
\end{table*}

\subsection{Modified DOVER for System Combination}


To further enhance robustness,
we also fuse the outputs of multiple spekaer diarization systems
by using a novel method, called modified DOVER. 
While widely utilized in other tasks such as 
speaker recognition (e.g., \cite{brummer2007fusion}) and automatic speech recognition (e.g., \cite{fiscus1997post}),
system fusion has been rarely explored for speaker diarization. 
Recently, \cite{stolcke2019dover} proposed a voting-based algorithm, called DOVER, where 
the multiple speaker diarization hypotheses are aligned in an iterative manner, and
the speaker for each time step 
is estimated by the weighted-voting from all the hypotheses.
While DOVER achieved significant improvement in their experiment,
it
has a limitation that
 overlapping speech cannot be handled correctly.

To achieve system combination even for overlapping speech,
 we propose a modified DOVER algorithm.
 The modified DOVER works as followings.
 The example procedure with three hypotheses is also shown in Fig. \ref{fig:modified_dover}.
 \begin{enumerate}
     \item Define a ``root'' hypothesis.
     \item Align each hypothesis from a different speaker diarization system with the root hypothesis by finding the speaker permutation that maximizes the total duration of overlap with the root hypothesis.
     \item For each speaker in the root hypothesis, each aligned hypothesis votes a weight for each time region for that speaker. If the total sum of voting weights exceeds the threshold, take that time region in the merged hypothesis.
     Note that a different hypothesis may have a different weight for voting. 
 \end{enumerate}

In the modified DOVER algorithm, there are a few things that need to be taken care of.
 Firstly, if the number of speakers in a hypothesis is larger than that of the root hypothesis,  there will be a speaker who does not have corresponding speaker of the root hypothesis after alignment in Step 2.
 In such a case, we simply discard such a speaker as exemplified in Speaker 4' of Hypothesis $\beta$ in Fig. \ref{fig:modified_dover}.
 Because of this, the number of speakers in the merged hypothesis never exceeds that of the root hypothesis.
 Secondly, there could be multiple choices for the root hypothesis. For example, we could use the hypothesis from the best system,
or we could use the hypothesis that has maximum number of speakers. 
We could even use the fused hypothesis by other technique (such as the original DOVER) as the root hypothesis.

In our final system, we combined four hypothesis.
We used the hypothesis from the best system on development set as the root hypothesis.
The voting weight was set to 1.0 for the best system,
and 0.34 for the other three systems. Threshold was set to 1.0.
Note that this setting eventually corresponds to keep using the hypothesis of the best system while appending the region
where the results of remaining three systems were coincident.

\section{Results and Analysis}
\label{sec:results}

The diarization results of the proposed systems and the official baseline of the challenge are shown in Table~\ref{table:results}. 
The difference between systems 1-3 is in the speaker embedding extraction model. 
System 4 then introduced the CSS module to handle overlapped speech. However, due to the leakage, false alarm was increased while miss rate was reduced, and overall DER on development set was not significantly improved. System 5 was the same as system 4 except that the leakage filtering was additionally applied. The speaker clusters from system 3 was used as the centroids for the filtering. It was observed that this filtering significantly reduced the DER on the development set. The system 5 also achieved significantly better DER and JER on the test set. 

In system 6, the AHC stopping threshold and minimum speaker duration were 
changed according to the grid search on the development set. 
Compared to system 3, DER was slightly reduced while JER was significantly reduced from 23.97\% to 19.90\%. 
System 7 was the same as system 5 except that system 6 was used for filtering out the leakage. 

Finally, systems 1, 2, 3, and 7 were fused using the modified DOVER algorithm by setting the hypothesis of system 7 as the root hypothesis. Compared to the best single system 7, DER was slightly improved while JER was marginally degraded. 
When we compare the system 8 (4th submission) with system 5 (3rd submission), we observe a significant improvement on the DER only in the test set.
We also observe 
that the JER was significantly degraded in the test set while it was significantly improved in the development set.
Due to the submission limit\footnote{We were able to submit only 4 systems due to the submission deadline while we were allowed to submit up to 5 systems.}, 
we could not do further analysis on these differences.
Our best result of 6.23\% DER was ranked 1st at the VoxSRC Challenge 2020.


\section{Conclusion}
This paper described the Microsoft speaker diarization system for 
monaural multi-talker recordings in the wild.
We proposed the speaker diarization system consists of
the state-of-the-art components
such as Res2Net-based speaker embedding extractor, conformer-based speech separation system
with the leakage filtering, and the modified  DOVER for  the  system  fusion. 
We evaluated the proposed system with the data set provided by the VoxSRC challenge 2020,
and finally achieved 3.71\% and 6.23\% of DERs on development set and evaluation set,  respectively,
being ranked 1st at the VoxSRC challenge 2020.



\bibliographystyle{IEEEtran}
\bibliography{refs}

\end{document}